\input{aipcheck}

\documentclass[
    ,final            
  ]
  {aipproc}

\layoutstyle{6x9}

\def\Journal#1#2#3#4{{#1} {#2} (#3) #4}

\def\PLB{{Phys. Lett.}  B}
\def\PRL{Phys. Rev. Lett.}

\def\RMP{Rev. Mod. Phys.}

\begin{document}

\title{Future Spin Physics at JLab\\ 12 GeV and Beyond}


\author{Kees de Jager}{
  address={Jefferson Laboratory, 
        Newport News, VA 23606, USA}
}

\begin{abstract}
 The project to upgrade the CEBAF accelerator at  Jefferson 
Lab to 12 GeV is presented. Most of the research program supporting that 
upgrade, will require a highly polarized beam, as will be illustrated by a few selected examples. To carry out that research program will require an extensively upgraded instrumentation in two of the existing experimental halls and the addition of a fourth hall. The plans for a high-luminosity electron-ion collider are briefly discussed.
\end{abstract}

\maketitle


\section{Introduction}
The design parameters of the Continuous Electron Beam Accelerator 
Facility (CEBAF) at the Thomas Jefferson National Accelerator Facility 
(JLab) were defined over two decades ago. Since then our understanding of 
the behaviour of strongly interacting matter has evolved significantly, 
providing important new classes of experimental questions which can be 
optimally adressed by a CEBAF-type accelerator at higher energy. 
The original design of the facility, coupled to 
developments in superconducting RF technology, makes it feasible to 
triple the initial design 
value of CEBAF's beam energy to 12~GeV in a  cost-effective 
manner.

The research program with the 12 GeV upgrade will provide 
breakthroughs in two key areas: (1) mapping gluonic excitations of mesons and understanding the origin of quark confinement and (2) searches for physics beyond the Standard Model. The upgrade will also provide important advances in two additional areas: (3) a direct exploration of the quark-gluon 
structure of the nucleon
and (4) the physics of nuclei to understand the QCD basis for the nucleon-nucleon force and how nucleons and mesons arise as an approximation to the underlying quark-gluon structure.
An overview of the upgrade research program is given in its 
Conceptual Design Report\cite{CDR}.

Lattice QCD 
calculations\cite{bali}  have convincingly illustrated the linear quark-quark 
potential necessary for confinement. The quark and 
anti-quark in a meson are sources of color electric flux, which is trapped in a flux tube connecting the $q$ and $\overline{q}$. 
However, very little is still 
known about the direct excitation of that flux tube. The observation 
of such direct manifestations of gluonic degrees of freedom will 
provide understanding of confinement\cite{exotics}. The quantum 
numbers of the flux tube, added to those of a $q\bar{q}$ meson, can produce 
exotic hybrids with unique $J^{PC}$ quantum numbers. These excitations can be probed far more effectively with photons than with $\pi$- or 
K-mesons, because the quark spins are aligned in the virtual 
vector-meson component of the photon. For a full partial-wave analysis of 
such excitations linearly polarized photons are a requisite.  The GlueX research program will be focused on a definitive 
measurement of the spectrum of exotic hybrid mesons,  
expected in a mass range from 1 to 2.5 GeV/c$^{2}$. 

One of the more compelling new opportunities with the 12 GeV upgrade will be a highly accurate measurement of 
the weak charge of the electron, via the parity-violating asymmetry in electron-electron (M{\o}ller) scattering. The achievable accuracy of such a measurement provides sensitivity to electron substructure to a scale of nearly 
30 TeV. 
The measurement is also sensitive to the existence of new neutral gauge bosons in the range of 
1 to 2 TeV; such model-dependent limits are complementary to those to be achieved by measurements 
at the Large Hadron Collider. Furthermore, the measurement will severely constrain the viability 
of SUSY models which violate R-parity. 
The upgraded beam energy will also make possible accurate measurements of parity violation in deep-
inelastic scattering (PVDIS). On an isoscalar target at moderate $x$ PVDIS is also sensitive to 
$\sin^2(\theta _W)$, thus providing a very sensitive test of electro-weak theory. 
Examples of additional PVDIS measurements are 
the value of $d(x)/u(x)$ as $x \rightarrow 1$, the search for evidence of charge symmetry violation at the partonic level, and the characterization of novel higher-twist effects. The PVDIS program will require the use of a new large-acceptance spectrometer/detector package, that can in parallel be used for a broad program of exclusive and semi-inclusive reaction processes.

A main focus of the research program will be the 
Generalized Parton Distributions (GPD) through the study of exclusive 
processes at large momentum transfer. The GPDs can be considered as 
overlap integrals between different components of the hadronic wave 
function\cite{ji}, governed by the selection of the final state. Measurements of 
these GPDs will thus make it possible to map out quark and gluon wave functions.
The orbital angular momentum contribution 
to the nucleon spin can be directly accessed through GPDs. Factorization is an 
essential ingredient in the extraction of GPDs. For Deeply Virtual Compton Scattering 
(DVCS) scaling has been shown to be valid already at 6 GeV, but  for other processes this has still to be established experimentally. 

\begin{figure}[h]
   \includegraphics[height=15pc]{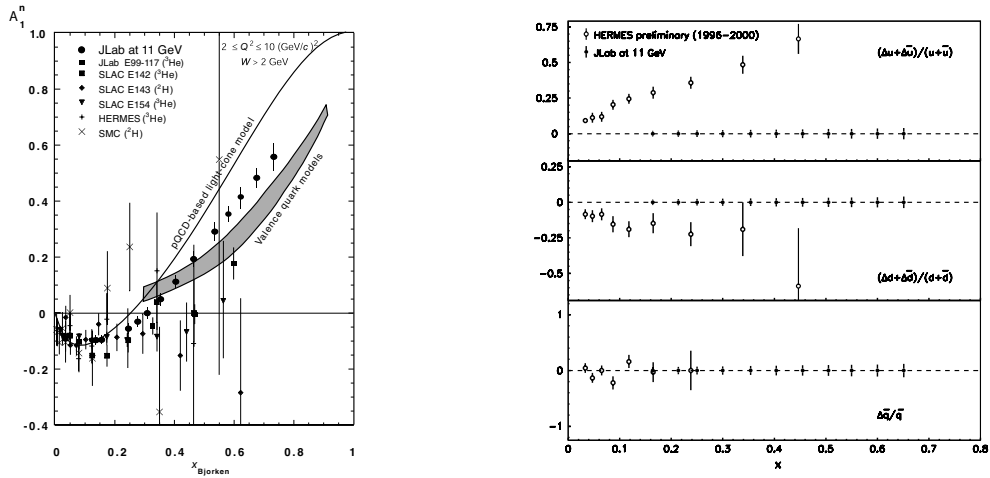}
   \caption{On the left is shown the projected measurement of $A_1^n$, on the right
   the projected determination of various combinations of polarized valence and sea 
   quark distributions from semi-inclusive deep inelastic scattering. } 
 \label{highx}
\end{figure}

One of the most fundamental properties of the nucleon is the structure of its quark distributions at higher $x$-values, where the 
physics of the valence quarks is cleanly exposed. 
The 12 GeV Upgrade will for the first time (by providing the necessary combination of high 
beam intensity and reach in $Q^2$) allow to map out the valence quark distributions at large 
$x$ with high precision. Most dynamical models predict that in 
the limit where a single valence up or down quark carries all of the momentum of the nucleon ($x \rightarrow 1$), 
it will also carry all of the spin polarization. Recent data from Hall A for the first time
show a hint of a possible upturn in the neutron
polarization asymmetry $A_1^n$ at an $x$-value of $\sim$0.6. Figure \ref{highx} (left)
 shows how $A_1^n$ can be measured up to $x$-values close to 0.8
outside the nucleon resonance region with the 12 GeV upgrade. 

\begin{figure}[h]
   \includegraphics[height=11pc]{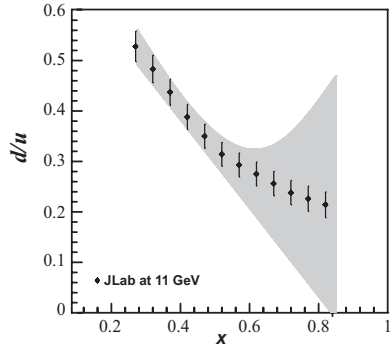}
   \caption{Projected measurement of the ratio of $d$- and $u$-quark momentum distributions, 
$d(x)/u(x)$, at large $x$. The shaded band represents the 
uncertainty in existing measurements due to nuclear Fermi motion effects. } 
 \label{doveru}
\end{figure}

There is a similar lack of data on 
other deep inelastic scattering observables in this region. 
One example 
is the ratio of down to up quarks in the proton, $d(x)/u(x)$, whose large-$x$ behavior is intimately 
related to the fact that the proton and neutron are the stable building blocks 
of nuclei. This ratio requires measurement of the structure 
function of the neutron as well as of the proton. Information about the neutron has so far been extracted from inclusive deuterium data, where it is difficult to disentangle from nuclear effects at large $x$. Figure \ref{doveru} shows 
the precision with which this fundamental ratio can be measured with the 12 GeV Upgrade. The 
proposed experiment will utilize a novel technique; detection of 
the slowly recoiling proton spectator will $tag$ scattering events on a nearly on-shell neutron in a 
deuteron target. An independent measurement of $d(x)/u(x)$ can be made by exploiting the mirror symmetry of $A$ = 3 nuclei in simultaneous measurements with $^3$He and $^3$H targets. Both methods are designed to largely eliminate the nuclear corrections, thereby permitting the $d/u$ 
ratio to be extracted with unprecedented precision. 

The precise way in which the spin of the nucleon is distributed 
among its quark and gluon constituents is one of the most fundamental questions that can be 
addressed in nonperturbative QCD. Most of the experiments so far have focused on measuring the total quark and 
gluon contribution to the nucleon spin in inclusive deep-inelastic scattering. In recent years the focus has moved to 
the investigation of specific aspects of the nucleon spin, such as the flavor asymmetries of sea quark 
distributions and quark transverse spin (transversity) distributions. 
The mapping of the flavor dependence of polarized valence and sea quark distributions and 
the determination of the quark transversity distributions require semi-inclusive measurements, in 
which the detected final-state hadron reveals information about the spin, flavor, and charge of 
the $struck$ quark participating in the deep-inelastic process. The 12 GeV Upgrade will provide 
a unique opportunity to perform semi-inclusive measurements with high precision over a wide 
kinematic range, producing a detailed picture of the spin structure of the nucleon. Figure \ref{highx} (right) shows
how polarized valence and sea
quark distributions can be extracted from semi-inclusive deep inelastic scattering
by detecting the leading $\pi^\pm$ and $K^\pm$ hadrons.

At 12 GeV, the details of the nucleon-nucleon force can be probed at distance scales much less than 
the pion Compton wave length, where the effects of two-pion exchange, vector-meson exchange, 
and quark exchange all compete. Although well-constrained phenomenologically by the large body of pp and np 
elastic scattering data, it is not yet understood under what circumstances the effective 
nuclear force can be described in terms of the exchange of mesons, and when it is more efficient to 
describe the force in terms of the underlying quark-gluon exchange forces. 
Alternatively, the atomic nucleus can be used as a laboratory to study how the underlying QCD non-Abelian 
degrees of freedom manifest themselves. The idea is here to strike 
a quark inside the nucleus with such velocity that one can uniquely witness how hadrons emerge  on their path through the 
nucleus. Our present sketchy understanding of this process will be vastly improved with the 12-GeV program at JLab.

\section{Accelerator}

At present CEBAF accelerates electrons to 6 GeV by recirculating the beam 
four times through two superconducting linacs, each producing an energy 
gain of 600 MeV per pass. 
Both linac tunnels provide sufficient space to install five additional 
newly designed cryomodules. 
The new cryomodules will each provide over 100 MV (compared to the 28 MV 
from the existing ones), by increasing the gradient to 20 MV/m and the number of cavity cells from five to seven. This will result in
a maximum energy gain per pass of 2.2 GeV, providing 
a maximum beam energy to the existing Halls A, B and C of 11 GeV. The new Hall D will be provided with the desired maximum energy of 12 GeV by adding a tenth 
arc and recirculating the beam a fifth time through one linac. A total of 90 
$\mu$A of CW beam can be provided at the maximum beam energy. Further 
modifications required are changing the dipoles in the  arcs 
from C-type to H-type magnets, replacing a large number of power 
supplies and doubling the capacity of the central helium liquifier .

\section{Existing Experimental Halls}

The CEBAF Large Acceptance Spectrometer (CLAS) 
in Hall B is used for experiments that require the detection of 
several, loosely correlated particles in the hadronic final state 
at a limited luminosity. The CLAS12 detector has evolved from CLAS to meet the 
basic requirements for the study of the structure of nucleons and nuclei with the CEBAF 12 GeV upgrade. The main features are: 
1) an operating luminosity of $L >$ 10$^{35}$ cm$^{-2}$s$^{-1}$ for hydrogen targets, a ten-fold increase over current CLAS operating conditions;
2) detection capabilities and particle identification for forward-going high momentum charged and neutral particles;
3) improved hermeticity for the detection of charged particles and photons. 
CLAS12 makes use of several existing detector components. Major new components include new superconducting torus coils that cover only the forward-angle range, a new gas \v Cerenkov counter 
for pion identification, additions to the electromagnetic calorimeters, and the central detector.

The Hall C facility has generally been used for experiments which 
require high luminosity at moderate resolution. The core 
spectrometers are the High Momentum Spectrometer (HMS) and the Short 
Orbit Spectrometer (SOS). The HMS has a maximum momentum of 7.6 GeV/c.
At a 12-GeV Jefferson Lab, the SOS spectrometer will be replaced by a new magnetic spectrometer, the Super High Momentum Spectrometer 
(SHMS), powerful enough to analyze charged particles with momenta approaching that of the highest energy beam.  
Charged particles with such high momenta are boosted by relativistic kinematics into the 
forward detection hemisphere. Therefore, the SHMS is designed through the use of a small horizontal bend magnet to achieve angles down to 
5.5$^o$ (and up to 25$^o$). The SHMS will cover a solid angle up to 5 msr, and boasts a large momentum and 
target acceptance. The magnetic spectrometer pair will be rigidly connected to a central pivot.

The present base instrumentation in Hall A has been used for experiments which require high luminosity and high 
resolution in momentum and/or angle of at least one of the reaction 
products. The central elements are the two High Resolution 
Spectrometers (HRS), to which recently a third spectrometer has been added with a large acceptance (BigBite). 
The beamline into Hall A will be upgraded so that the hall will be able to accept the full range of beam energies available for two major purposes. 
The first will be to continue the use of the three existing spectrometers. The second purpose for Hall A will be to stage major installation experiments. With a diameter of over 50 m, Hall A is 
the largest experimental hall at Jefferson Lab and  can easily accommodate major installations such as the proposed parity-violation setups. 

\section{Hall D}

The GlueX experiment will be housed in a new aboveground experimental hall (Hall D) located at the east end of the CEBAF north linac. A collimated 
beam of linearly polarized photons (with 40\% polarization) of energy 8.5 to 9 GeV, optimum for the production of exotic hybrids in its 
expected mass range, will be produced 
via coherent bremsstrahlung with 12 GeV electrons. This requires carefully aligned thin diamond crystal radiators. The scattered electron from the bremsstrahlung will be tagged with 
sufficient precision to determine the photon energy to within 0.1\%. 

The GlueX detector uses an existing 2.25 T superconducting solenoid that is 
currently being refurbished. An existing 3000-element lead-glass electromagnetic calorimeter will 
be reconfigured to match the downstream aperture of the solenoid. Inside the full length of the solenoid, a lead and scintillating 
fiber electromagnetic calorimeter will provide position and energy measurement for photons and 
TOF information for charged particles. A simple start counter will surround the 30 cm long liquid hydrogen target. This in turn will be surrounded by cylindrical straw-tube drift-chambers which 
will fill the region between the target and the cylindrical calorimeter. Planar drift chambers will 
be placed inside the solenoid downstream of the target to provide accurate track reconstruction for 
charged particles going in the forward direction. 

This detector configuration has 4$\pi$
hermeticity and momentum/energy and position information for charged particles and photons produced from incoming 9 GeV 
photons. It has 
been carefully optimized to carry out partial wave analysis of many-particle final states.  
The final planned photon flux is $10^{8}$ photons/s. At this 
 flux the experiment will accumulate in one year of running a factor of 
 100 more meson data  than are presently available even from pion production.

\section{Summary}

In April of 2004 the US Department of Energy (DOE) signed CD-0 approval for the 12 GeV Upgrade project, acknowledging the mission need for this project. Then, in February of 2006 DOE approved the preliminary baseline range through the second Critical Decision. Two further review processes in increasing level of detail, spaced 12 to 18 months apart, have to be successfully passed before construction funding for the project will be allocated. The 12 GeV Upgrade project is the only large construction project underway at the Nuclear Physics program office in the DOE Office of Science.  At present 11 GeV beam to at least one hall and the start of the first experiments is expected in 2012, with full operations in all four halls to commence by late 2014.

At Jefferson Lab a plan for the next upgrade, involving an electron-ion collider  is already being developed. Initial design studies have yielded a promising concept for up to 150 GeV protons colliding with up to 7 GeV electrons (or positrons) at a luminosity of close to $10^{35}$ cm$^{-2}$s$^{-1}$, thus at a center-of-mass energy of up to 65 GeV. Both electrons or positrons and protons would be circulating in two storage rings with four interaction regions at maximum polarization for either beam. Light ions up to mass $\sim$ 40 would also be available.

\begin{theacknowledgments}

This work was supported by DOE contract DE-AC05-06OR23177, under which the Jefferson Science Associates, LLC, (JSA) operates the Thomas Jefferson National Accelerator Facility. 
\end{theacknowledgments}

\end{document}